\documentclass[amsmath,amssymb,amsbsy,prl,twocolumn,tightenlines,superscriptaddress,floatfix]{revtex4-2}
\usepackage{amsfonts} 
\usepackage{amsmath} 
\usepackage{amssymb}
\usepackage{physics}
\usepackage{graphicx,nicefrac}
\usepackage{blindtext}
\usepackage{mathtools,newtxmath}
\usepackage{lineno}
\usepackage{cancel}
\usepackage{scalerel}
\usepackage[normalem]{ulem}
\usepackage{xfrac}
\usepackage{bm}
\usepackage[usenames, dvipsnames]{color} 
\usepackage{dcolumn}
\usepackage{epic} 
\usepackage{epsfig}
\usepackage{eucal} 
\usepackage{esdiff}
\usepackage{graphicx}
\usepackage[breaklinks,colorlinks = true,linkcolor = magenta,urlcolor=magenta,citecolor=red]{hyperref}
\usepackage{mathrsfs}
\usepackage{soul}
\usepackage{subfigure}
\usepackage{textcomp}
\pagecolor{white}
\usepackage{verbatim}
\usepackage{wrapfig} 
\usepackage{xy} 
\usepackage{lineno}
\usepackage{float}
\usepackage{import}

\newcommand{\landau}{ Landau Institute for Theoretical Physics, RAS, 142432, Chernogolovka, Semenova 1A, Moscow region, 
Russia} 
\newcommand{\HSE}{NRU Higher School of Economics, 101000 Myasnitskaya 20, Moscow, Russia}
\newcommand{\kyoto}{Department of Physics, Kyoto University, Kitashirakawa Oiwakecho Sakyoku Kyoto, 606-8502, Japan}
\newcommand{\iitk}{Department of Mechanical Engineering, Indian Institute of Technology Kanpur, Kanpur 208016, India}
\begin{document}
\title{Geometric Intermittency in Turbulence}
\begin{abstract} 

Equal-time scaling exponents in fully developed turbulence typically exhibit
	non anomalous scaling in the inverse cascade of two-dimensional (2D)
	turbulence and anomalous scaling in three dimensions. We demonstrate
	that multiscaling is not confined to longitudinal, scalar velocity
	increments, but also emerges in increments associated with the
	magnitude and orientation of the velocity vector. This decomposition
	uncovers a  multiscaling in the 2D inverse cascade, which remains
	obscured when using conventional structure functions. Our results
	highlight a decoupling between velocity amplitude and flow geometry,
offering new insight into the statistical structure of turbulent cascades as
well as showing how different classes of multiscaling emerge.  

\end{abstract}
\author{Ritwik Mukherjee} 
\email{ritwik.mukherjee@icts.res.in}
\affiliation{International Centre for Theoretical Sciences, Tata Institute of
Fundamental Research, Bengaluru 560089, India} 
\author{Siddhartha Mukherjee}
\email{smukherjee@iitk.ac.in} 
\affiliation{\iitk} 
\author{I. V. Kolokolov}
\email{Igor.Kolokolov@gmail.com} 
\affiliation{\landau} 
\affiliation{\HSE}
\author{V. V. Lebedev} 
\email{lebede@itp.ac.ru} 
\affiliation{\landau}
\affiliation{\HSE} 
\author{Takeshi Matsumoto}
\email{takeshi@kyoryu.scphys.kyoto-u.ac.jp} 
\affiliation{\kyoto}
\author{Samriddhi Sankar Ray} 
\email{samriddhisankarray@gmail.com}
\affiliation{International Centre for Theoretical Sciences, Tata Institute of
Fundamental Research, Bengaluru 560089, India} 
\maketitle

Kolmogorov's 1941 theory remains the peg on which our understanding of the
statistics of  non-equilibrium stationary states of three-dimensional (3D)
fully developed, statistically homogeneous and isotropic turbulence rests~\cite{Frisch-Book}.
It provides a statistical physics framework to interpret universality in
turbulence through two-point correlation functions. Such $p$-th order
correlation functions  $S^{\rm L}_p (r) = \langle (\delta u^{\rm L}_r )^p
\rangle$ ---  called longitudinal (hence the superscript L) structure functions
in turbulence --- are defined through the scalar velocity increment $\delta
u^{\rm L}_r \equiv \delta {\bf u}_r \cdot {\hat {\bf r}}$, where the $\delta
{\bf u}_r$ is the (vector) velocity difference between points A and B separated
by a distance vector ${\bf r}$ and ${\hat {\bf r}} = {\bf r}/|{\bf r}|$.  The
angular brackets $\langle \cdot \cdot \cdot \rangle$ correspond to an
\textit{ensemble} average, typically over space and time in the
non-equilibrium, statistically, steady state. For separations within the
inertial range of turbulence $\eta \ll r \ll \ell$, where $\eta$ is the (small)
dissipative scale and $\ell$ the large (integral scale) of turbulence,
and where the flow is known, from experiments and numerical
simulations, to be statistically homogeneous and isotropic, scale-invariance leads to
$S^{\rm L}_p (r) \propto r^{\zeta_p^{\rm
L}}$~\cite{Frisch-Book}. The equal-time longitudinal exponents  $\zeta_p^{\rm
L} = p/3$ scale linearly with $p$ --- the normal scaling form --- follow from
assumptions of homogeneity, isotropy and a finite energy dissipation rate
$\varepsilon$ (and hence flux), along with an $\ell$ independence, in the limit
of high Reynolds number. 

While such a normal scaling form for the third-order exponent
yields $\zeta_3^{\rm L} = 1$, a manifestation of a
scale-independent energy flux and consistent with the Karman-Howarth
relation~\cite{Frisch-Book},  we know that the measured exponents deviate
significantly from the $p/3$ scaling for $p \neq 3$.  In particular, in the
inertial range, $S^{\rm L}_p(r) = C^{\rm L}_p \left (\frac{\ell}{r}\right
)^{p/3 - \zeta^{\rm L}_p}\left (\varepsilon r \right)
^{p/3}$~\cite{Frisch-Book}, where $\zeta_p^{\rm L}$ is a convex, monotonically
increasing function of $p$ while satisfying the constraint $\zeta_3^{\rm L} =
1$~\cite{Frisch-Book}. This anomalous, multiscaling, rationalised through the
Parisi-Frisch multifractal formalism~\cite{Frisch-Parisi}, is a fundamental
property of turbulence whose origins lie in the ubiquitous intermittent nature
of turbulent flows.

\begin{figure}
\includegraphics[width = 1.00\linewidth]{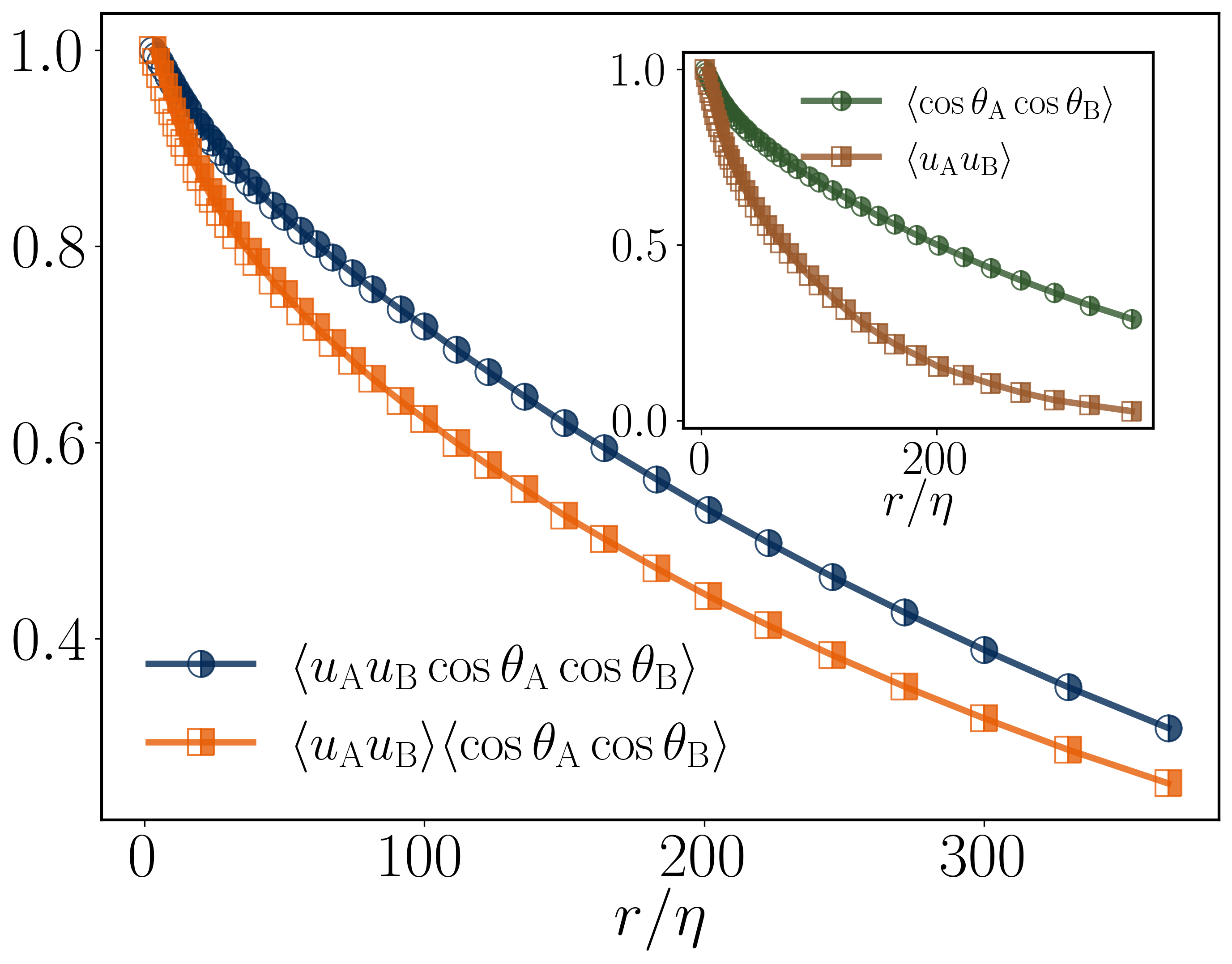}
	\caption{A comparison of the mixed correlator $\langle 2u_{\rm A}u_{\rm B}\cos\theta_{\rm A}\cos\theta_{\rm B} \rangle$ and 
	$\langle 2u_{\rm A}u_{\rm B}\rangle \langle\cos\theta_{\rm A}\cos\theta_{\rm B} \rangle$ as well as (inset)
	$\langle \cos \theta_{\rm A} \cos \theta_{\rm B} \rangle $ and $\langle u_{\rm A} u_{\rm B} \rangle$,  normalised to 1 at $r = 0$, 
	in three-dimensional turbulence.}
	\label{fig:cross-term3d}
\end{figure}

Intriguingly buried in the definition of the scalar increment $\delta u^{\rm
L}_r$ are the tangled contributions coming from the statistics of the velocity
amplitudes $u_{\rm A}$ and $u_{\rm B}$ as well as the cosines of the angles ---
$\cos \theta_{\rm A} = {\bf \hat{u}}_{\rm A}\cdot {\bf \hat{r}}$ and $ \cos
\theta_{\rm B} = {\bf \hat{u}}_{\rm B}\cdot {\bf \hat{r}}$ --- the unit
velocity vectors make with the unit vector between points A and B: For example,
$S^{\rm L}_2 (r) = \langle u_{\rm A}^2\cos^2 \theta_{\rm A} + u_{\rm B}^2\cos^2
\theta_{\rm B} - 2u_{\rm A}u_{\rm B}\cos\theta_{\rm A}\cos\theta_{\rm B}
\rangle$.  We confirm from numerical simulations that $u_{\rm A}^2$ and $\cos^2
\theta_{\rm A}$ are uncorrelated (likewise for point B) and hence $\langle
u_{\rm A}^2\cos^2 \theta_{\rm A} \rangle = \langle u_{\rm A}^2\rangle
\langle\cos^2 \theta_{\rm A}\rangle = c_0 \, v^2_{\rm rms}$, where $ c_0 =
\langle\cos^2 \theta_{\rm A}\rangle = \nicefrac{1}{3}$ follows trivially from a
uniform distribution of orientations in 3D. But what about the 
correlations involving the velocity amplitudes and the cosine of the angles?
Indeed, can the mixed correlation functions be simply written as products of
correlations between the amplitudes and the angles?

Data from Direct Numerical Simulations
(DNSs), whose details are discussed later, show unambiguously (see
Fig.~\ref{fig:cross-term3d}) that $\langle u_{\rm A}u_{\rm B}\cos\theta_{\rm
A}\cos\theta_{\rm B} \rangle \neq \langle u_{\rm A}u_{\rm B}\rangle
\langle\cos\theta_{\rm A}\cos\theta_{\rm B} \rangle$. Furthermore, the velocity amplitude
correlation $\langle u_{\rm A}u_{\rm B}\rangle$ decays faster with $r$ than the correlation
function of the angular part $\langle\cos\theta_{\rm A}\cos\theta_{\rm B} \rangle$  
(Fig.~\ref{fig:cross-term3d}, inset). This result~\cite{footnote1} underlines
the fact that the statistics of the amplitudes and geometry of the velocity
vectors are just as complex and perhaps equally fundamental, which deserves a
thorough investigation. 

There is one other reason which motivates the present study. Recent
measurements~\cite{BuariaSreeniPRL23} of the scaling exponents ${\zeta_p^{\rm
T}}$ of the transverse structure functions --- $S^{\rm T}_p (r) = \langle \left
( \delta u^{\rm T}_r \right )^p \rangle$, where $\delta u^{\rm T}_r = \delta
{\bf u} \cdot {\bf \hat{r}^{{\rm T}}}$ with ${\bf \hat{r}} \cdot
{\bf\hat{r}^{{\rm T}}} = 0$  ---  find a systematic, statistically significant
departure from ${\zeta_p^{\rm L}}$ suggestive of the role of geometry. Specifically, Buaria and
Sreenivasan~\cite{BuariaSreeniPRL23} show that ${\zeta_p^{\rm T}}$ has stronger
signatures of multiscaling, with a possible saturation of ${\zeta_p^{\rm T}}$ 
for $p \gtrsim 8$.

This motivates the question of whether the classical projected increments, from
which longitudinal and transverse structure functions are built, truly capture
the strongest manifestations of intermittency --- or whether alternative increments
can reveal even richer multiscaling behaviour. By separating velocity
amplitudes from the local flow geometry via increments of $\cos\theta_{\rm A}$
and $\cos\theta_{\rm B}$, we uncover a family of structure functions that
exhibit equally strong anomalous scaling and, strikingly, reveal persistent
intermittency in the \emph{nominally non-intermittent} inverse-cascade regime
of two-dimensional turbulence when viewed solely through $\zeta_p^{\rm L}$.

We use the publicly available data from the Johns Hopkins Turbulence Database
(JHTD)~\cite{perlman2007data,li2008public,yeung2012dissipation} --- with
Taylor-scale based Reynolds numbers $Re_\lambda = 418$  and $610 $ corresponding to resolutions 
$1024^3$ and $4096^3$, respectively--- which is obtained
from pseudo-spectral direct numerical simulations (DNSs) of the incompressible,
3D Navier-Stokes equation with forcing at large scales. The exponents we report (in the main text 
and the Appendix) are, within error-bars, the same for both these Reynolds number; the data we present 
in the figures are from the larger Reynolds number simulation.

We define structure functions $S^{\rm u}_p \equiv \langle
|\delta u_r |^p \rangle = C_p^{\rm u} \left (\frac{\ell}{r}\right )^{p/3 -
\xi_p}\left (\varepsilon r \right) ^{p/3}$, for the amplitude increment $\delta
u_r = u_{\rm B} - u_{\rm A}$, and $S^{\delta \cos\theta}_p \equiv \langle
|\delta \cos \theta_r |^p \rangle =  C_p^{\delta \cos \theta}  \left
(\frac{\ell}{r}\right )^{-\chi_p}$, for the cosine-angle increment $\delta \cos
\theta_r =  \cos \theta_{\rm B} - \cos \theta_{\rm A}$. The
scaling form Ansatz is in direct corollary with that assumed for the
longitudinal structure functions in the inertial range. In Appendix A, we show
consistency of such scaling forms and find, in particular, the Kolmogorov
constants $C_2^{\rm L} \approx C_2^{\rm u} \gg C_2^{\delta \cos \theta}$~\cite{Ni2013}.

\begin{figure}
	\includegraphics[width = 1.00\linewidth]{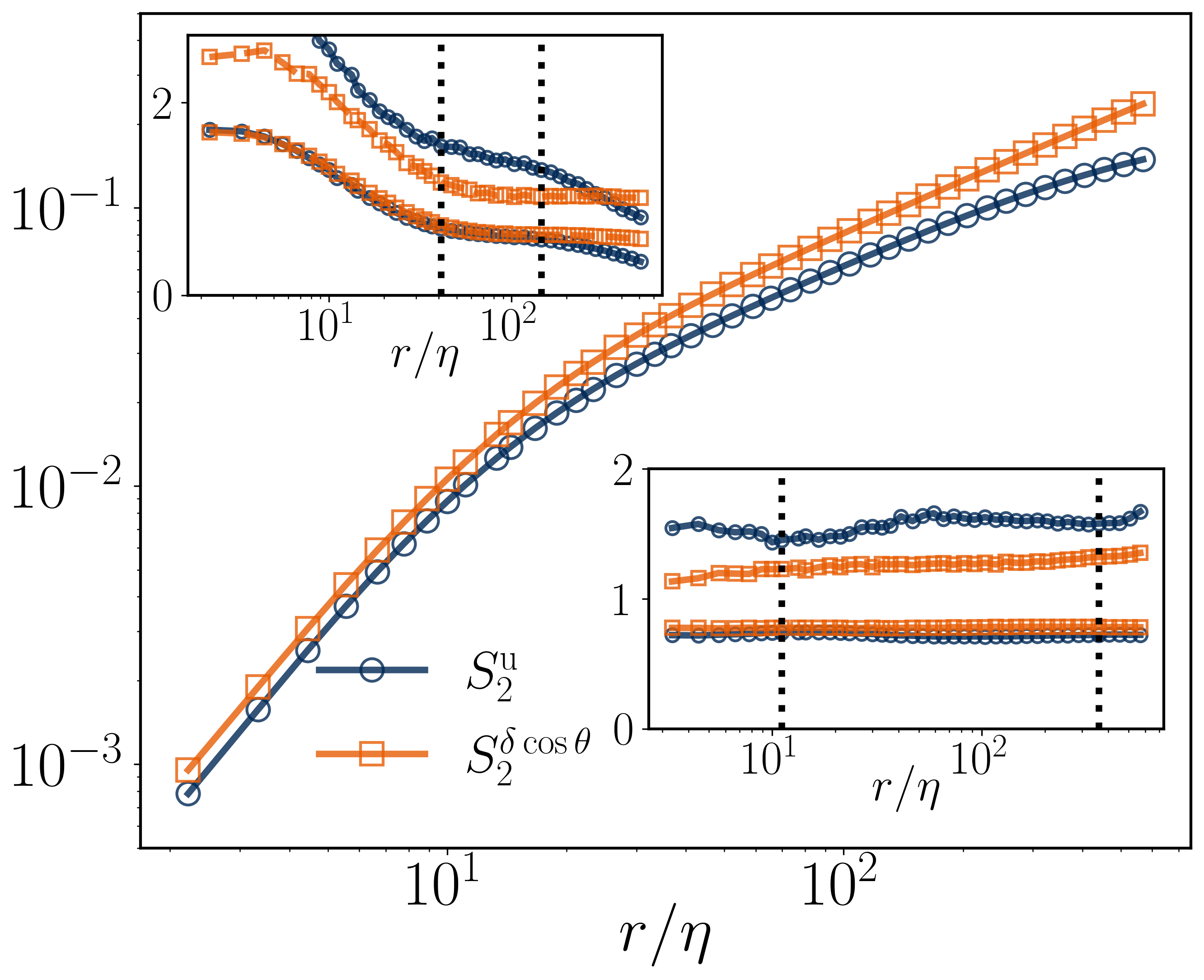}
	\caption{Log-log plots of the second-order structure functions vs $r/\eta$. 
	Upper inset: Local slopes of  $S^{\rm u}_p$ and $S^{\delta \cos \theta}_p$ for $p = 2$ (lower set of curves) and $p=6$ 
	(upper set of curves); the pair of vertical lines denote the inertial range. 
	Lower inset: Analogous plots as those in the upper inset but for local slopes 
	extracted via ESS.}
	\label{fig:S2}
\end{figure}

In Fig.~\ref{fig:S2} we show a representative log-log plots of $S^{\rm u}_2$
and $S^{\delta \cos \theta}_2$ against $r/\eta$ (as is common) from which emerges the
scaling exponents $\xi_2$ and $\chi_2$.  From purely dimensional arguments it
is reasonable to expect $\xi_2 = \zeta_2^{\rm L}$. In the upper inset of
Fig.~\ref{fig:S2} (lower pair of curves) we show a semilog plot of the local
slopes for $S^{\rm u}_2$ and $S^{\delta \cos \theta}_2$. In the inertial range, denoted by
the pair of vertical, dashed lines, we find a plateau, whose value yields
$\xi_2 \approx \chi_2 \approx \zeta_2^{\rm L} = \nicefrac{2}{3}$ (up to
intermittency corrections).

What is most revealing is the behaviour of higher-order exponents. For
illustration, we consider the sixth-order structure functions. A local-slope
analysis (upper inset of Fig.~\ref{fig:S2}) gives the corresponding exponents.
While the velocity-amplitude increment shows only a marginal inertial-range
plateau, the plateau for $S^{\delta \cos\theta}_6$ is markedly clearer.
Moreover, the resulting exponents differ substantially from the classical
longitudinal value (dash-dotted line), with $\chi_6 \neq \xi_6 < \zeta_6^{\rm
L}$. Because high-order exponents in turbulence are notoriously sensitive to
finite-Reynolds-number effects~\cite{Frisch-Book}, we further validate this result using extended
self-similarity (ESS)~\cite{BenziESS,RayESS}. The ESS local slopes, shown in
the lower inset of Fig.~\ref{fig:S2} for $p = 2$ and $p = 6$, confirm that the
exponent ratios $\tilde{\chi}_p = \chi_p/\chi_3$ and $\tilde{\xi}_p =
\xi_p/\xi_3$ deviate significantly from the corresponding longitudinal ratios
$\zeta_p^{\rm L}/\zeta_3^{\rm L}$ throughout the inertial range. This
persistent discrepancy underscores the distinct and stronger intermittency
encoded in the geometric increments.

\begin{figure}
	\includegraphics[width = 1.00\linewidth]{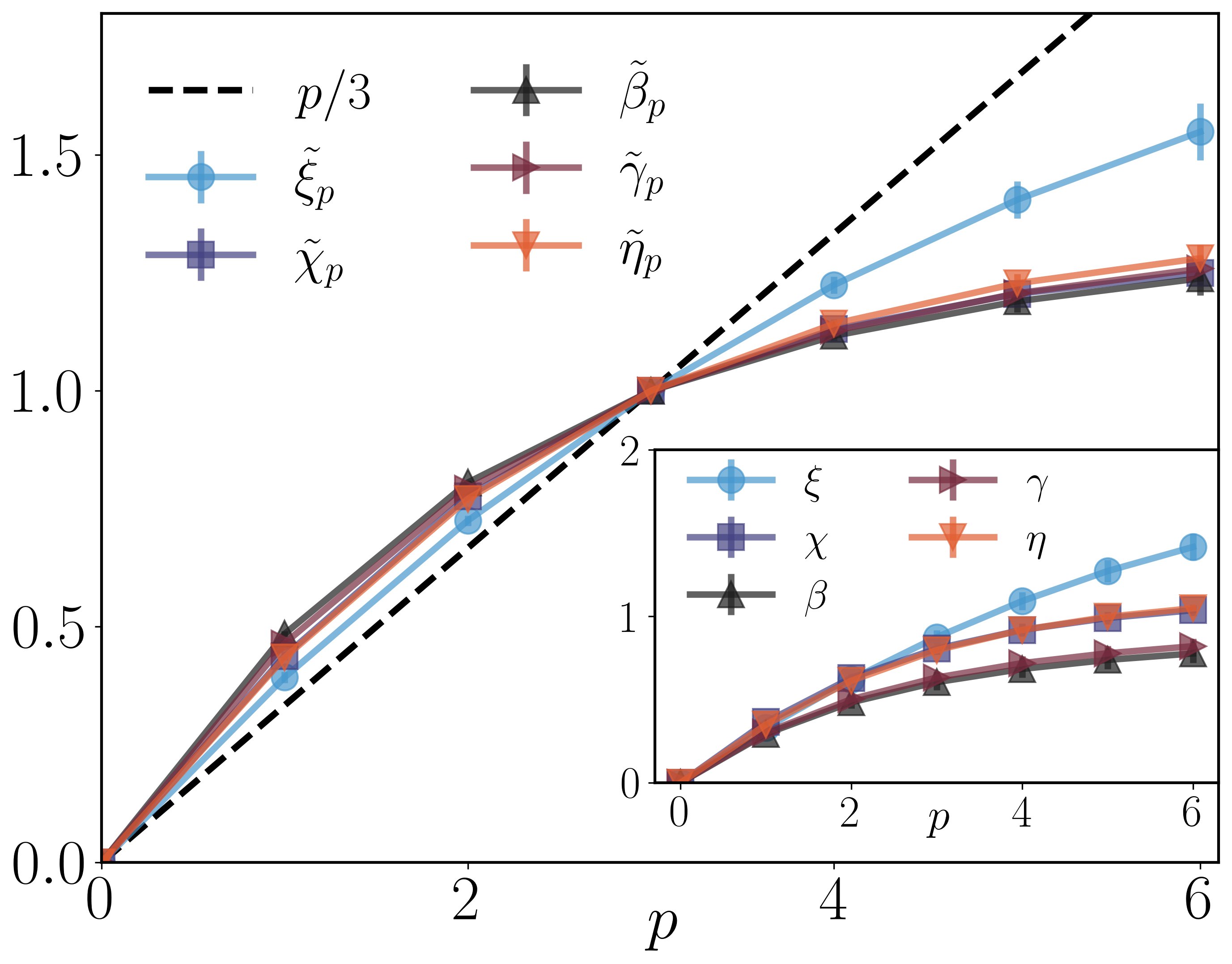}
	\caption{Plots of the scaling exponent ratios $\tilde{\xi}_p$, $\tilde{\chi}_p$, $\tilde{\beta}_p$, $\tilde{\gamma}_p$ and $\tilde{\eta}_p$ (obtained via ESS) 
	and (inset) the bare exponents $\xi_p$, $\chi_p$, $\beta_p$, $\gamma_p$ and $\eta_p$ for three-dimensional turbulence.}
	\label{fig:3d_zetaP}
\end{figure}

A note of caution is warranted. Because our observables involve velocity
amplitudes and directional cosines, there is no K\'arm\'an--Howarth--type
constraint that would enforce a unit third-order exponent. Consistently, as
shown in the inset of Fig.~\ref{fig:3d_zetaP} and in Table~\ref{tab:exponents},
we find $\chi_3 = \xi_3 \neq 1$ within error bars. Therefore, our use of ESS
with the third-order structure function is strictly procedural: It extends the
apparent scaling range (lower inset of Fig.~\ref{fig:S2}) and yields exponent
ratios rather than bare exponents~\cite{BenziESS,RayESS}. Throughout this manuscript we use the term
``bare exponents'' to denote those obtained from direct log--log fits of the
structure functions versus $r$, in contrast to the ESS-based exponent ratios.

Nevertheless, the comparison of the local slopes of the second and sixth order moments ---
and the fact that they seem to deviate from normal scaling even more strongly from the classical
longitudinal exponent --- is our cue to calculate these new set of scaling exponents as well as the 
exponent ratios through ESS.

In Fig.~\ref{fig:3d_zetaP} we show plots of the exponent ratios $\tilde{\chi}_p$ and
$\tilde{\xi}_p$ versus $p$~\cite{footnote2}, with both showing a remarkable
degree of multiscaling.  To confirm that this is not an artefact of extended
self similarity, we calculate the bare exponents as a function of $p$ as well.
In the inset of Fig.~\ref{fig:3d_zetaP} we show a plot of these scaling
exponents vs $p$ and find an equally compelling case of multiscaling with the
same contrasting trends between the scaling exponents for $S^{\rm u}_p$ and
$S^{\delta \cos \theta}_p$ (see Table~\ref{tab:exponents}) as already noted for
the exponent ratios.

\begin{figure}
	\includegraphics[width = 1.00\linewidth]{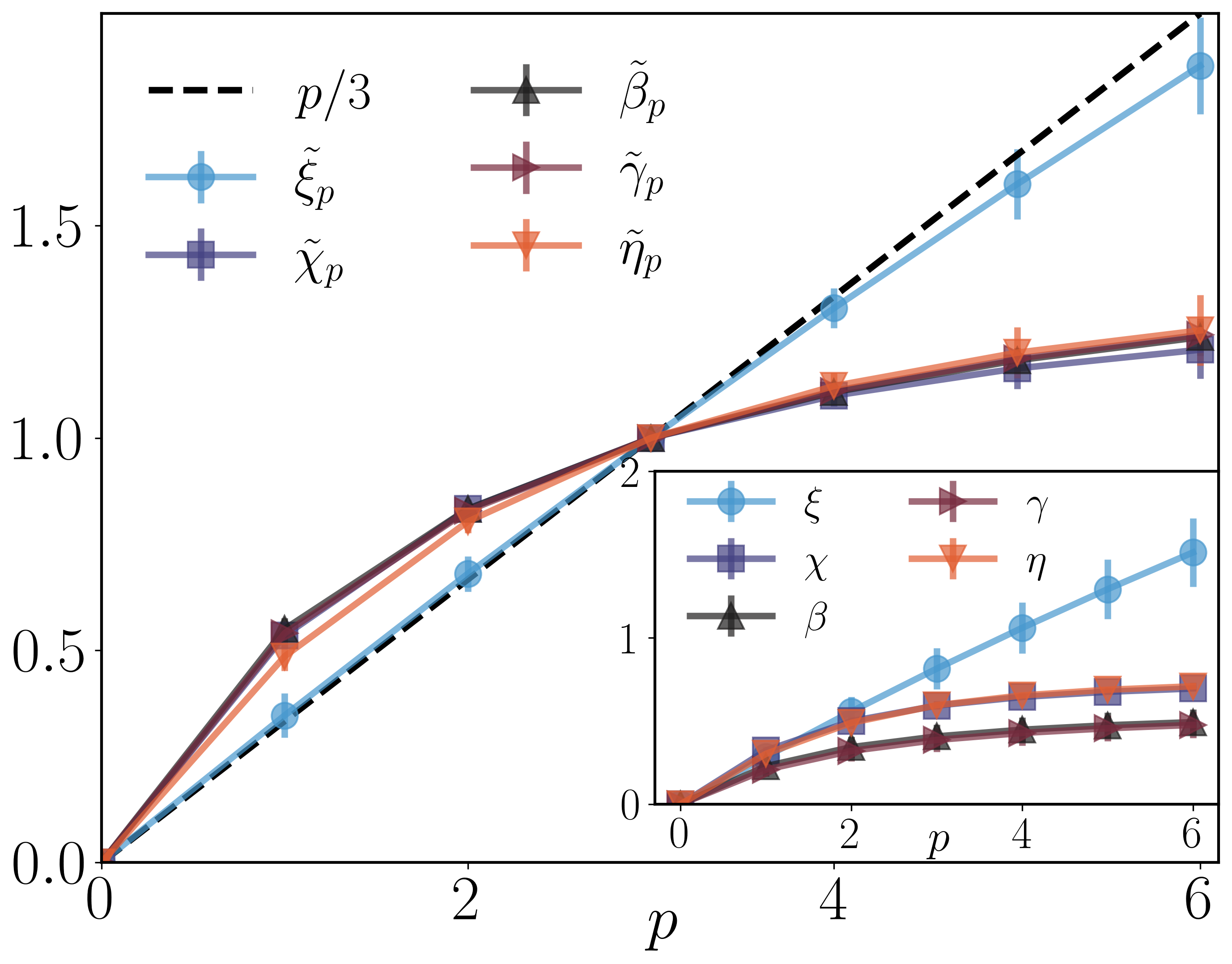}
	\caption{Plots of the scaling exponent ratios $\tilde{\chi}_p$, $\tilde{\xi}_p$, $\tilde{\beta}_p$, $\tilde{\gamma}_p$ and $\tilde{\eta}_p$ (obtained via ESS) and (inset) the bare exponents $\chi_p$, $\xi_p$, $\beta_p$, $\gamma_p$ and $\eta_p$ for two-dimensional turbulence.}
	\label{fig:2d_zetaP}
\end{figure}

\begin{table*}
	\centering
	\small
	\begin{tabular}{|c|c|c|c|c|c|c|c|c|c|}
	\hline
	$p$ & \multicolumn{5}{c|}{Three-dimensional Turbulence} & \multicolumn{4}{c|}{Two-dimensional Turbulence} \\
		\hline
		& $\xi_p$ & $\tilde{\xi}_p$ & $\chi_p$ & $\tilde{\chi}_p$ & $\zeta^L_p$ &  $\xi_p$ & $\tilde{\xi}_p$ & $\chi_p$ & $\tilde{\chi}_p$ \\
		\hline
		1 & $0.33 \pm 0.02$ & $0.39 \pm 0.01$ & $0.36 \pm 0.01$ & $0.44 \pm 0.01$ & $0.3777 \pm 0.0001$ &    $0.29 \pm 0.05$ & $0.35 \pm 0.05$ & $0.32 \pm 0.02$ & $0.53 \pm 0.03$ \\
		2 & $0.62 \pm 0.03$ & $0.72 \pm 0.01$ & $0.63 \pm 0.02$ & $0.78 \pm 0.00$ & $0.7091 \pm 0.0001$ &    $0.56 \pm 0.09$ & $0.68 \pm 0.04$ & $0.50 \pm 0.03$ & $0.83 \pm 0.02$ \\
		3 & $0.87 \pm 0.04$ & $1.00 $ 				& $0.80 \pm 0.03$ & $1.00 $ 			& $1.0059 \pm 0.0002$ &    $0.81 \pm 0.12$ & $1.00$ & $0.59 \pm 0.02$ & $1.00$ \\
		4 & $1.09 \pm 0.05$ & $1.22 \pm 0.02$ & $0.91 \pm 0.03$ & $1.13 \pm 0.01$ & $1.2762 \pm 0.0002$ &   $1.06 \pm 0.15$ & $1.31 \pm 0.05$ & $0.64 \pm 0.01$ & $1.10 \pm 0.03$ \\
		5 & $1.27 \pm 0.07$ & $1.40 \pm 0.04$ & $0.98 \pm 0.02$ & $1.21 \pm 0.02$ & $1.5254 \pm 0.0005$ &   $1.29 \pm 0.18$ & $1.60 \pm 0.08$ & $0.68 \pm 0.01$ & $1.16 \pm 0.05$ \\
		6 & $1.42 \pm 0.08$ & $1.55 \pm 0.06$ & $1.03 \pm 0.02$ & $1.25 \pm 0.03$ & $1.757 \pm 0.001$ &       $1.51 \pm 0.20$ & $1.88 \pm 0.11$ & $0.70 \pm 0.01$ & $1.21 \pm 0.07$ \\
		\hline
	\end{tabular}
	\caption{A summary of the various equal-time scaling exponents and the exponent ratios for both 3D and 2D turbulence for 
		$1 \le p \le 6$. For comparison, the equal-time longitudinal exponents $\zeta_p^{\rm L}$ are given from the same 
		simulations and consistent with those reported earlier from DNSs~\cite{Iyer2017} (or indeed in shell 
		models~\cite{MitraPRL2004}) for three-dimensional turbulence has been listed in column 6.}
	\label{tab:exponents}
\end{table*}

While our investigation of the increments $\delta\cos\theta$ and $u$ is
motivated by the structure of the longitudinal increment, the results reveal a
broader and more fundamental feature of intermittency that is encoded in the
geometry of the flow. In Appendix~B we show that several other angle-based
increments, such as $\delta\theta_r = \theta_{\rm A}-\theta_{\rm B}$ (with
exponents $\eta_p$ and $-\pi \leq \delta\theta_r \le \pi$),
$\delta\sin\theta_r = \sin\theta_{\rm A}-\sin\theta_{\rm B}$ (with exponents
$\beta_p$), and $\sin\delta\theta_r = \sin[\theta_{\rm A}-\theta_{\rm B}]$
(with exponents $\gamma_p$), also exhibit clear signatures of multiscaling with
$\chi_p \approx \eta_p \neq \beta_p \approx \gamma_p$.  Given the lack of a
K\'arm\'an--Howarth constraint, this could be because such structure functions
are (a) perhaps more susceptible to sub-dominant, finite-Re effects and (b) the
ones defined through sinusoidal functions range differently than those defined
via cosine or just the increment $\delta \theta$.  However, the corresponding
ESS exponent ratios --- which compensate for the unknown infrared and
ultraviolet corrections that contaminate the finite-Re inertial
range~\cite{BenziESS,RayESS} --- collapse onto a single curve. This collapse
demonstrates that all angular increments probe the same underlying singular
structure associated with the geometry of the velocity field. Taken together
with the results in Table~\ref{tab:add_exponents_3D} (and Table~\ref{tab:add_exponents_2D}) in Appendix B, 
this establishes a new
universality class of multiscaling exponents tied to the geometric properties
of turbulence.

A natural question is how these new exponents relate to the traditional
multiscaling exponents $\zeta_p^{\rm L}$ and $\zeta_p^{\rm T}$ of the
longitudinal and transverse structure functions.  Since the latter arise from
projections of the velocity increment onto the separation vector, it is useful
to express the inverse transformation that reconstructs the velocity magnitude
increment from its projected components and the local geometry.

Inverting the standard projection yields
\begin{equation}
\begin{pmatrix}
u_{\rm B} \\
u_{\rm A}
\end{pmatrix}
=
\frac{1}{\sin(\theta_{\rm B}-\theta_{\rm A})}
\begin{pmatrix}
-\sin\theta_{\rm A} & \cos\theta_{\rm A} \\
-\sin\theta_{\rm B} & \cos\theta_{\rm B}
\end{pmatrix}
\begin{pmatrix}
\delta u_r^{\rm L} \\
\delta u_r^{\rm T}
\end{pmatrix}
\end{equation}
and hence
\begin{equation}
\delta u_r = 
\frac{\delta u_r^{\rm L}\,\delta\sin\theta_r - \delta u_r^{\rm T}\,\delta\cos\theta_r}
{\sin\delta\theta_r},
\label{eq:bridge}
\end{equation}
with $\delta\sin\theta_r = \sin\theta_{\rm B}-\sin\theta_{\rm A}$ and 
$\sin\delta\theta_r = \sin(\theta_{\rm B}-\theta_{\rm A})$.

Assuming typical H\"older behaviour  
$\delta u_r \sim r^{h_u}$, $\delta u_r^{\rm L} \sim r^{h^{\rm L}_u}$, 
$\delta u_r^{\rm T}\sim r^{h^{\rm T}_u}$, 
$\delta\sin\theta_r \sim r^{h_\theta^{\rm s}}$, 
$\delta\cos\theta_r \sim r^{h_\theta^{\rm c}}$, 
and $\sin\delta\theta_r \sim r^{h_\theta}$,  
Eq.~\eqref{eq:bridge} gives
$h_u = \min\!\left[\,h_u^{\rm L}+h_\theta^{\rm s}-h_\theta,\, h_u^{\rm T}+h_\theta^{\rm c}-h_\theta\,\right]$.

For $p=1$, these H\"older exponents coincide with the measured scaling exponents, so that  
$h_u^{\rm L}\sim\zeta_1^{\rm L}$, $h_u^{\rm T}\sim\zeta_1^{\rm T}$, 
$h_\theta^{\rm s}\sim\beta_1$, $h_\theta^{\rm c}\sim\chi_1$, and $h_\theta\sim\gamma_1$ (Appendix~B).  
This yields the exact relation
$\xi_1 = \zeta_1^{\rm L} + \beta_1 - \gamma_1$,
which we verify numerically.  
For $p>1$, mixed correlations enter Eq.~\eqref{eq:bridge}, and the simple H\"older connection 
no longer holds, consistent with multifractal intermittency.

Taken together, this bridge relation and the collapse of the ESS exponents demonstrate that the 
velocity–direction field carries a distinct and universal multiscaling structure. Conventional 
longitudinal and transverse structure functions therefore capture only part of the intermittency; 
the geometric degrees of freedom contain an equally fundamental, and in the case of the 2D inverse cascade, 
previously hidden, signature of turbulent multiscaling.

All of this  throws up a tantalising question in the context of two-dimensional
turbulence~\cite{boffetta2012}.  We know that the inverse cascade regime $L \gg
r \gg L_{\rm F}$, where $L_{\rm F}$ is the forcing scale and $L$ the
large scale of the system set by the drag coefficient, in forced, two-dimensional turbulence obeys the
Kolmogorov normal scaling  theory and the associated longitudinal exponents
$\zeta_p^{\rm L} = p/3$~\cite{tsang2005,perlekar2009,ssray2011,Pandit2D2017}.
It is tempting to investigate if structure functions constructed as before from
velocity magnitudes and cosines of angles show any evidence of anomalous
scaling for their associated exponents. 

We perform pseudo-spectral DNSs of the incompressible, 2D Navier-Stokes
equation in the vorticity-streamfunction formulation in a $2\pi$ periodic
domain with hyperviscosity ($\nu \Delta^8$)  and hypodrag ($\alpha
\Delta^{-1}$) terms. We force the system at high wavenumber $\mathbf{k}$ such
that $k_f-1 < |\mathbf{k}| < k_f + 1$ with $k_f=997$. Hyperviscosity $\nu =
1.05  \times 10^{-47}$ and hypodrag $\alpha = 35$~\cite{Takeshi2022}. From
statistically steady velocity field obtained from our DNSs, we measure, like in
the 3D problem discussed above, the equal-time scaling exponents $\chi_p$,
$\xi_p$, $\beta_p$, $\gamma_p$, and $\eta_p$ along with their counterparts
$\tilde{\chi}_p$, $\tilde{\xi}_p$, $\tilde{\beta}_p$, $\tilde{\gamma}_p$, and $\tilde{\eta}_p$ 
obtained from ESS~\cite{BenziESS,RayESS}).

In the inset of Fig.~\ref{fig:2d_zetaP} we plot the bare exponents
 as a function of the order $p$.
Remarkably, the exponents 
for the structure function constructed from angles show strong anomalous
scaling with a suggestion that the exponents saturate at large $p$ in a manner
more pronounced than what is seen in 3D turbulence (see
Fig.~\ref{fig:3d_zetaP}). In the main panel of Fig.~\ref{fig:2d_zetaP} we show
the corresponding exponent ratios. While $\tilde{\xi}_p$ does not really deviate 
from the normal scaling, a hitherto undetected multiscaling emerges 
for the exponents related to the angles. This is very similar to what we have seen 
for three-dimensional turbulence, including the curious observation that while
$\chi_p \approx \eta_p \neq \beta_p \approx \gamma_p$, the exponent ratios collapse once 
more on to a single, multiscaling curve.

These results suggest that the fundamental
property of full-developed turbulence --- intermittency --- manifests itself
not just in the projected velocity increments but in the more basic quantities
of velocity amplitudes and its geometry. This is a significant departure from our
conventional, accepted wisdom of working with longitudinal and transverse
structure functions as \textit{the} probe to investigate multiscaling. This in turn, 
uncovers a way to detect an incipient
multiscaling --- already suggested in a recent work by M\"uller and
Krstulovic~\cite{MullerPRF} --- in the inverse cascade regime of
two-dimensional turbulence which is masked in the more conventional route via
longitudinal structure functions. This is perhaps the most striking 
outcome of the new class of equal-time exponents, summarised in Table~\ref{tab:exponents}, 
that we report. Such multiscaling is of course accompanied by non-Gaussian tails in the probability 
distribution functions (PDFs) of the increments themselves as we show in Appendix B of this manuscript.

Let us make a final remark on why the geometric measures show a
much larger degree of multiscaling than the ones associated with the velocity
amplitudes or indeed the conventional, longitudinal structure functions. By
assuming similar scaling forms for longitudinal and transverse structure
functions, Eq.~\eqref{eq:bridge}, the Parisi-Frisch formalism leads to $\xi_p =
\inf\limits_{h^{\rm L}_u} \{p[h^{\rm L}_u + g(h^{\rm L}_u)] + d - D(h_u^{\rm
L})\}$, where $d$ is the space dimension and we assume $\frac{\delta \sin
\theta_r}{\sin \delta \theta_r} \propto r^{g(h^{\rm L}_u)}$ in
Eq.~\eqref{eq:bridge}. We now make the Ansatz that $g(h^{\rm L}_u) = -\lambda
h^{\rm L}_u$ which yields $\xi_p = \zeta^{\rm L}_{p(1+\lambda)}$, where
$\lambda$ is a fitting parameter.  Numerically, we find that such
a relation between the two sets of exponents seem consistent for $\lambda =
-0.15$. Such a conjecture, consistent with the multifractal formalism, would
indeed lead to the conclusion that the geometry of the flow is more
intermittent --- with stronger multiscaling in the associated exponents ---
than what is observed for longitudinal structure functions.

In summary, we have shown that intermittency in turbulence is not exhausted by
longitudinal and transverse velocity increments: Geometric increments of the
velocity field display equally strong, and in some regimes stronger,
multiscaling. This reveals a previously hidden intermittency in the 2D inverse
cascade and identifies a universal class of geometric scaling exponents.
This, of course, leads to questions of
whether long-lived vortical structures play a more significant role in making
flows intermittent, especially in 2D, than appreciated hitherto. Our results also suggest that the
geometry of turbulent velocity fields plays a fundamental role in cascade
dynamics~\cite{MukherjeeLocal} and opens a route to probing intermittency,
blind to conventional structure functions, in other
flows~\cite{Ray-Irreversible}. 

\begin{acknowledgments}

RM, SM and SSR acknowledges several insightful discussions with Jason Picardo.  RM
	acknowledges the Infosys-TIFR Leading Edge Travel Grant 2025 which
	facilated part of this research.  SM acknowledges the IITK Initiation
	Grant project IITK/ME/2024316 and support from the Govt. of India Grant
	ANRF/ECRG/2024/002467/ENS.  IVK acknowledges support from the Ministry
	of Science and Higher Education of the Russian Federation. The work of
	VVL  was supported by Grant No.  23-72-30006 of Russian Science
	Foundation.  TM was Grants-in-Aid for Scientific Research KAKENHI No.
	(C) 23K03247 from JSPS. SSR gratefully acknowledges the hospitality of
	the Landau Institute for Theoretical Physics where this work was
	initiated during an Indo-Russian meeting and the  Infosys International
	Exchange Program which facilitated this visit.  SSR acknowledges  the Indo–French Centre for the Promotion of Advanced Scientific Research (IFCPAR/CEFIPRA, project no. 6704-1) for support.  This research was supported in
	part by the International Centre for Theoretical Sciences (ICTS) for
	the program --- 10th Indian Statistical Physics Community Meeting
	(code: ICTS/10thISPCM2025/04). The simulations were performed on the
	ICTS clusters Mario, Tetris, and Contra. RM and SSR acknowledge the
	support of the DAE, Government of India, under projects nos.
	12-R\&D-TFR-5.10-1100 and RTI4001. 

\end{acknowledgments}

\bibliographystyle{apsrev4-2} 
 \bibliography{references}

\begin{thebibliography}{22}%
\makeatletter
\providecommand \@ifxundefined [1]{%
 \@ifx{#1\undefined}
}%
\providecommand \@ifnum [1]{%
 \ifnum #1\expandafter \@firstoftwo
 \else \expandafter \@secondoftwo
 \fi
}%
\providecommand \@ifx [1]{%
 \ifx #1\expandafter \@firstoftwo
 \else \expandafter \@secondoftwo
 \fi
}%
\providecommand \natexlab [1]{#1}%
\providecommand \enquote  [1]{``#1''}%
\providecommand \bibnamefont  [1]{#1}%
\providecommand \bibfnamefont [1]{#1}%
\providecommand \citenamefont [1]{#1}%
\providecommand \href@noop [0]{\@secondoftwo}%
\providecommand \href [0]{\begingroup \@sanitize@url \@href}%
\providecommand \@href[1]{\@@startlink{#1}\@@href}%
\providecommand \@@href[1]{\endgroup#1\@@endlink}%
\providecommand \@sanitize@url [0]{\catcode `\\12\catcode `\$12\catcode
  `\&12\catcode `\#12\catcode `\^12\catcode `\_12\catcode `\%12\relax}%
\providecommand \@@startlink[1]{}%
\providecommand \@@endlink[0]{}%
\providecommand \url  [0]{\begingroup\@sanitize@url \@url }%
\providecommand \@url [1]{\endgroup\@href {#1}{\urlprefix }}%
\providecommand \urlprefix  [0]{URL }%
\providecommand \Eprint [0]{\href }%
\providecommand \doibase [0]{https://doi.org/}%
\providecommand \selectlanguage [0]{\@gobble}%
\providecommand \bibinfo  [0]{\@secondoftwo}%
\providecommand \bibfield  [0]{\@secondoftwo}%
\providecommand \translation [1]{[#1]}%
\providecommand \BibitemOpen [0]{}%
\providecommand \bibitemStop [0]{}%
\providecommand \bibitemNoStop [0]{.\EOS\space}%
\providecommand \EOS [0]{\spacefactor3000\relax}%
\providecommand \BibitemShut  [1]{\csname bibitem#1\endcsname}%
\let\auto@bib@innerbib\@empty
\bibitem [{\citenamefont {Frisch}(1996)}]{Frisch-Book}%
  \BibitemOpen
  \bibfield  {author} {\bibinfo {author} {\bibfnamefont {U.}~\bibnamefont
  {Frisch}},\ }\href@noop {} {\emph {\bibinfo {title} {Turbulence: The Legacy
  of A. N. Kolmogorov}}}\ (\bibinfo  {publisher} {Cambridge University Press},\
  \bibinfo {year} {1996})\BibitemShut {NoStop}%
\bibitem [{\citenamefont {Parisi}\ and\ \citenamefont
  {Frisch}(1985)}]{Frisch-Parisi}%
  \BibitemOpen
  \bibfield  {author} {\bibinfo {author} {\bibfnamefont {G.}~\bibnamefont
  {Parisi}}\ and\ \bibinfo {author} {\bibfnamefont {U.}~\bibnamefont
  {Frisch}},\ }\href {https://cir.nii.ac.jp/crid/1570291225411264384}
  {\bibfield  {journal} {\bibinfo  {journal} {Proceedings of the International
  School of Physics Enrico Fermi, Course LXXXVIII, Varenna, 1985}\ } (\bibinfo
  {year} {1985})}\BibitemShut {NoStop}%
\bibitem [{foo({\natexlab{a}})}]{footnote1}%
  \BibitemOpen
  \href@noop {} {}\bibinfo {howpublished} {We have confirmed that a
  reconstruction of the second-order structure function by using the various
  terms in its expansion reproduces identically the longitudinal structure
  functions constructed from the velocity increments.} (\bibinfo {year}
  {{}}{\natexlab{a}})\BibitemShut {NoStop}%
\bibitem [{\citenamefont {Buaria}\ and\ \citenamefont
  {Sreenivasan}(2023)}]{BuariaSreeniPRL23}%
  \BibitemOpen
  \bibfield  {author} {\bibinfo {author} {\bibfnamefont {D.}~\bibnamefont
  {Buaria}}\ and\ \bibinfo {author} {\bibfnamefont {K.~R.}\ \bibnamefont
  {Sreenivasan}},\ }\href {https://doi.org/10.1103/PhysRevLett.131.204001}
  {\bibfield  {journal} {\bibinfo  {journal} {Phys. Rev. Lett.}\ }\textbf
  {\bibinfo {volume} {131}},\ \bibinfo {pages} {204001} (\bibinfo {year}
  {2023})}\BibitemShut {NoStop}%
\bibitem [{\citenamefont {Perlman}\ \emph {et~al.}(2007)\citenamefont
  {Perlman}, \citenamefont {Burns}, \citenamefont {Li},\ and\ \citenamefont
  {Meneveau}}]{perlman2007data}%
  \BibitemOpen
  \bibfield  {author} {\bibinfo {author} {\bibfnamefont {E.}~\bibnamefont
  {Perlman}}, \bibinfo {author} {\bibfnamefont {R.}~\bibnamefont {Burns}},
  \bibinfo {author} {\bibfnamefont {Y.}~\bibnamefont {Li}},\ and\ \bibinfo
  {author} {\bibfnamefont {C.}~\bibnamefont {Meneveau}},\ }in\ \href
  {https://doi.org/https://doi.org/10.1145/1362622.1362654} {\emph {\bibinfo
  {booktitle} {Proceedings of the 2007 ACM/IEEE Conference on
  Supercomputing}}}\ (\bibinfo {year} {2007})\ pp.\ \bibinfo {pages}
  {1--11}\BibitemShut {NoStop}%
\bibitem [{\citenamefont {Li}\ \emph {et~al.}(2008)\citenamefont {Li},
  \citenamefont {Perlman}, \citenamefont {Wan}, \citenamefont {Yang},
  \citenamefont {Meneveau}, \citenamefont {Burns}, \citenamefont {Chen},\ and\
  \citenamefont {Eyink}}]{li2008public}%
  \BibitemOpen
  \bibfield  {author} {\bibinfo {author} {\bibfnamefont {Y.}~\bibnamefont
  {Li}}, \bibinfo {author} {\bibfnamefont {E.}~\bibnamefont {Perlman}},
  \bibinfo {author} {\bibfnamefont {M.}~\bibnamefont {Wan}}, \bibinfo {author}
  {\bibfnamefont {Y.}~\bibnamefont {Yang}}, \bibinfo {author} {\bibfnamefont
  {C.}~\bibnamefont {Meneveau}}, \bibinfo {author} {\bibfnamefont
  {R.}~\bibnamefont {Burns}}, \bibinfo {author} {\bibfnamefont
  {A.}~\bibnamefont {Chen}, \bibfnamefont {Shiyi an d~Szalay}},\ and\ \bibinfo
  {author} {\bibfnamefont {G.}~\bibnamefont {Eyink}},\ }\href
  {https://doi.org/https://doi.org/10.1080/14685240802376389} {\bibfield
  {journal} {\bibinfo  {journal} {Journal of Turbulence}\ ,\ \bibinfo {pages}
  {N31}} (\bibinfo {year} {2008})}\BibitemShut {NoStop}%
\bibitem [{\citenamefont {Yeung}\ \emph {et~al.}(2012)\citenamefont {Yeung},
  \citenamefont {Donzis},\ and\ \citenamefont
  {Sreenivasan}}]{yeung2012dissipation}%
  \BibitemOpen
  \bibfield  {author} {\bibinfo {author} {\bibfnamefont {P.}~\bibnamefont
  {Yeung}}, \bibinfo {author} {\bibfnamefont {D.}~\bibnamefont {Donzis}},\ and\
  \bibinfo {author} {\bibfnamefont {K.}~\bibnamefont {Sreenivasan}},\ }\href
  {https://doi.org/https://doi.org/10.1017/jfm.2012.5} {\bibfield  {journal}
  {\bibinfo  {journal} {Journal of Fluid Mechanics}\ }\textbf {\bibinfo
  {volume} {700}},\ \bibinfo {pages} {5} (\bibinfo {year} {2012})}\BibitemShut
  {NoStop}%
\bibitem [{\citenamefont {Ni}\ and\ \citenamefont {Xia}(2013)}]{Ni2013}%
  \BibitemOpen
  \bibfield  {author} {\bibinfo {author} {\bibfnamefont {R.}~\bibnamefont
  {Ni}}\ and\ \bibinfo {author} {\bibfnamefont {K.-Q.}\ \bibnamefont {Xia}},\
  }\href {https://doi.org/10.1103/PhysRevE.87.023002} {\bibfield  {journal}
  {\bibinfo  {journal} {Phys. Rev. E}\ }\textbf {\bibinfo {volume} {87}},\
  \bibinfo {pages} {023002} (\bibinfo {year} {2013})}\BibitemShut {NoStop}%
\bibitem [{\citenamefont {Benzi}\ \emph {et~al.}(1993)\citenamefont {Benzi},
  \citenamefont {Ciliberto}, \citenamefont {Tripiccione}, \citenamefont
  {Baudet}, \citenamefont {Massaioli},\ and\ \citenamefont {Succi}}]{BenziESS}%
  \BibitemOpen
  \bibfield  {author} {\bibinfo {author} {\bibfnamefont {R.}~\bibnamefont
  {Benzi}}, \bibinfo {author} {\bibfnamefont {S.}~\bibnamefont {Ciliberto}},
  \bibinfo {author} {\bibfnamefont {R.}~\bibnamefont {Tripiccione}}, \bibinfo
  {author} {\bibfnamefont {C.}~\bibnamefont {Baudet}}, \bibinfo {author}
  {\bibfnamefont {F.}~\bibnamefont {Massaioli}},\ and\ \bibinfo {author}
  {\bibfnamefont {S.}~\bibnamefont {Succi}},\ }\href
  {https://doi.org/10.1103/PhysRevE.48.R29} {\bibfield  {journal} {\bibinfo
  {journal} {Phys. Rev. E}\ }\textbf {\bibinfo {volume} {48}},\ \bibinfo
  {pages} {R29} (\bibinfo {year} {1993})}\BibitemShut {NoStop}%
\bibitem [{\citenamefont {Chakraborty}\ \emph {et~al.}(2010)\citenamefont
  {Chakraborty}, \citenamefont {Frisch},\ and\ \citenamefont {Ray}}]{RayESS}%
  \BibitemOpen
  \bibfield  {author} {\bibinfo {author} {\bibfnamefont {S.}~\bibnamefont
  {Chakraborty}}, \bibinfo {author} {\bibfnamefont {U.}~\bibnamefont
  {Frisch}},\ and\ \bibinfo {author} {\bibfnamefont {S.~S.}\ \bibnamefont
  {Ray}},\ }\href {https://doi.org/10.1017/S0022112010000595} {\bibfield
  {journal} {\bibinfo  {journal} {Journal of Fluid Mechanics}\ }\textbf
  {\bibinfo {volume} {649}},\ \bibinfo {pages} {275–285} (\bibinfo {year}
  {2010})}\BibitemShut {NoStop}%
\bibitem [{foo({\natexlab{b}})}]{footnote2}%
  \BibitemOpen
  \href@noop {} {}\bibinfo {howpublished} {The exponents are calculated as the
  mean height of the plateaus from a local slope analysis (Fig.~\ref{fig:S2},
  insets) and the standard deviation of these are our error-bars.}
  ({\natexlab{b}})\BibitemShut {NoStop}%
\bibitem [{\citenamefont {Iyer}\ \emph {et~al.}(2017)\citenamefont {Iyer},
  \citenamefont {Sreenivasan},\ and\ \citenamefont {Yeung}}]{Iyer2017}%
  \BibitemOpen
  \bibfield  {author} {\bibinfo {author} {\bibfnamefont {K.~P.}\ \bibnamefont
  {Iyer}}, \bibinfo {author} {\bibfnamefont {K.~R.}\ \bibnamefont
  {Sreenivasan}},\ and\ \bibinfo {author} {\bibfnamefont {P.~K.}\ \bibnamefont
  {Yeung}},\ }\href {https://doi.org/10.1103/PhysRevE.95.021101} {\bibfield
  {journal} {\bibinfo  {journal} {Phys. Rev. E}\ }\textbf {\bibinfo {volume}
  {95}},\ \bibinfo {pages} {021101} (\bibinfo {year} {2017})}\BibitemShut
  {NoStop}%
\bibitem [{\citenamefont {Mitra}\ and\ \citenamefont
  {Pandit}(2004)}]{MitraPRL2004}%
  \BibitemOpen
  \bibfield  {author} {\bibinfo {author} {\bibfnamefont {D.}~\bibnamefont
  {Mitra}}\ and\ \bibinfo {author} {\bibfnamefont {R.}~\bibnamefont {Pandit}},\
  }\href {https://doi.org/10.1103/PhysRevLett.93.024501} {\bibfield  {journal}
  {\bibinfo  {journal} {Phys. Rev. Lett.}\ }\textbf {\bibinfo {volume} {93}},\
  \bibinfo {pages} {024501} (\bibinfo {year} {2004})}\BibitemShut {NoStop}%
\bibitem [{\citenamefont {Boffetta}\ and\ \citenamefont
  {Ecke}(2012)}]{boffetta2012}%
  \BibitemOpen
  \bibfield  {author} {\bibinfo {author} {\bibfnamefont {G.}~\bibnamefont
  {Boffetta}}\ and\ \bibinfo {author} {\bibfnamefont {R.~E.}\ \bibnamefont
  {Ecke}},\ }\href
  {https://doi.org/https://doi.org/10.1146/annurev-fluid-120710-101240}
  {\bibfield  {journal} {\bibinfo  {journal} {Annual review of fluid
  mechanics}\ }\textbf {\bibinfo {volume} {44}},\ \bibinfo {pages} {427}
  (\bibinfo {year} {2012})}\BibitemShut {NoStop}%
\bibitem [{\citenamefont {Tsang}\ \emph {et~al.}(2005)\citenamefont {Tsang},
  \citenamefont {Ott}, \citenamefont {Antonsen~Jr},\ and\ \citenamefont
  {Guzdar}}]{tsang2005}%
  \BibitemOpen
  \bibfield  {author} {\bibinfo {author} {\bibfnamefont {Y.-K.}\ \bibnamefont
  {Tsang}}, \bibinfo {author} {\bibfnamefont {E.}~\bibnamefont {Ott}}, \bibinfo
  {author} {\bibfnamefont {T.~M.}\ \bibnamefont {Antonsen~Jr}},\ and\ \bibinfo
  {author} {\bibfnamefont {P.~N.}\ \bibnamefont {Guzdar}},\ }\href
  {https://doi.org/10.1103/PhysRevE.71.066313} {\bibfield  {journal} {\bibinfo
  {journal} {Phys. Rev. E}\ }\textbf {\bibinfo {volume} {71}},\ \bibinfo
  {pages} {066313} (\bibinfo {year} {2005})}\BibitemShut {NoStop}%
\bibitem [{\citenamefont {Perlekar}\ and\ \citenamefont
  {Pandit}(2009)}]{perlekar2009}%
  \BibitemOpen
  \bibfield  {author} {\bibinfo {author} {\bibfnamefont {P.}~\bibnamefont
  {Perlekar}}\ and\ \bibinfo {author} {\bibfnamefont {R.}~\bibnamefont
  {Pandit}},\ }\href {https://doi.org/10.1088/1367-2630/11/7/073003} {\bibfield
   {journal} {\bibinfo  {journal} {New Journal of Physics}\ }\textbf {\bibinfo
  {volume} {11}},\ \bibinfo {pages} {073003} (\bibinfo {year}
  {2009})}\BibitemShut {NoStop}%
\bibitem [{\citenamefont {Ray}\ \emph {et~al.}(2011)\citenamefont {Ray},
  \citenamefont {Mitra}, \citenamefont {Perlekar},\ and\ \citenamefont
  {Pandit}}]{ssray2011}%
  \BibitemOpen
  \bibfield  {author} {\bibinfo {author} {\bibfnamefont {S.~S.}\ \bibnamefont
  {Ray}}, \bibinfo {author} {\bibfnamefont {D.}~\bibnamefont {Mitra}}, \bibinfo
  {author} {\bibfnamefont {P.}~\bibnamefont {Perlekar}},\ and\ \bibinfo
  {author} {\bibfnamefont {R.}~\bibnamefont {Pandit}},\ }\href
  {https://doi.org/10.1103/PhysRevLett.107.184503} {\bibfield  {journal}
  {\bibinfo  {journal} {Physical Review Letters}\ }\textbf {\bibinfo {volume}
  {107}},\ \bibinfo {pages} {184503} (\bibinfo {year} {2011})}\BibitemShut
  {NoStop}%
\bibitem [{\citenamefont {Pandit}\ \emph {et~al.}(2017)\citenamefont {Pandit},
  \citenamefont {Banerjee}, \citenamefont {Bhatnagar}, \citenamefont {Brachet},
  \citenamefont {Gupta}, \citenamefont {Mitra}, \citenamefont {Pal},
  \citenamefont {Perlekar}, \citenamefont {Ray}, \citenamefont {Shukla},\ and\
  \citenamefont {Vincenzi}}]{Pandit2D2017}%
  \BibitemOpen
  \bibfield  {author} {\bibinfo {author} {\bibfnamefont {R.}~\bibnamefont
  {Pandit}}, \bibinfo {author} {\bibfnamefont {D.}~\bibnamefont {Banerjee}},
  \bibinfo {author} {\bibfnamefont {A.}~\bibnamefont {Bhatnagar}}, \bibinfo
  {author} {\bibfnamefont {M.}~\bibnamefont {Brachet}}, \bibinfo {author}
  {\bibfnamefont {A.}~\bibnamefont {Gupta}}, \bibinfo {author} {\bibfnamefont
  {D.}~\bibnamefont {Mitra}}, \bibinfo {author} {\bibfnamefont
  {N.}~\bibnamefont {Pal}}, \bibinfo {author} {\bibfnamefont {P.}~\bibnamefont
  {Perlekar}}, \bibinfo {author} {\bibfnamefont {S.~S.}\ \bibnamefont {Ray}},
  \bibinfo {author} {\bibfnamefont {V.}~\bibnamefont {Shukla}},\ and\ \bibinfo
  {author} {\bibfnamefont {D.}~\bibnamefont {Vincenzi}},\ }\href
  {https://doi.org/10.1063/1.4986802} {\bibfield  {journal} {\bibinfo
  {journal} {Physics of Fluids}\ }\textbf {\bibinfo {volume} {29}},\ \bibinfo
  {pages} {111112} (\bibinfo {year} {2017})}\BibitemShut {NoStop}%
\bibitem [{\citenamefont {Kishi}\ \emph {et~al.}(2022)\citenamefont {Kishi},
  \citenamefont {Matsumoto},\ and\ \citenamefont {Toh}}]{Takeshi2022}%
  \BibitemOpen
  \bibfield  {author} {\bibinfo {author} {\bibfnamefont {T.}~\bibnamefont
  {Kishi}}, \bibinfo {author} {\bibfnamefont {T.}~\bibnamefont {Matsumoto}},\
  and\ \bibinfo {author} {\bibfnamefont {S.}~\bibnamefont {Toh}},\ }\href
  {https://doi.org/10.1103/PhysRevFluids.7.064604} {\bibfield  {journal}
  {\bibinfo  {journal} {Phys. Rev. Fluids}\ }\textbf {\bibinfo {volume} {7}},\
  \bibinfo {pages} {064604} (\bibinfo {year} {2022})}\BibitemShut {NoStop}%
\bibitem [{\citenamefont {M\"uller}\ and\ \citenamefont
  {Krstulovic}(2025)}]{MullerPRF}%
  \BibitemOpen
  \bibfield  {author} {\bibinfo {author} {\bibfnamefont {N.~P.}\ \bibnamefont
  {M\"uller}}\ and\ \bibinfo {author} {\bibfnamefont {G.}~\bibnamefont
  {Krstulovic}},\ }\href {https://doi.org/10.1103/PhysRevFluids.10.L012601}
  {\bibfield  {journal} {\bibinfo  {journal} {Phys. Rev. Fluids}\ }\textbf
  {\bibinfo {volume} {10}},\ \bibinfo {pages} {L012601} (\bibinfo {year}
  {2025})}\BibitemShut {NoStop}%
\bibitem [{\citenamefont {Mukherjee}\ \emph {et~al.}(2024)\citenamefont
  {Mukherjee}, \citenamefont {Murugan}, \citenamefont {Mukherjee},\ and\
  \citenamefont {Ray}}]{MukherjeeLocal}%
  \BibitemOpen
  \bibfield  {author} {\bibinfo {author} {\bibfnamefont {S.}~\bibnamefont
  {Mukherjee}}, \bibinfo {author} {\bibfnamefont {S.~D.}\ \bibnamefont
  {Murugan}}, \bibinfo {author} {\bibfnamefont {R.}~\bibnamefont {Mukherjee}},\
  and\ \bibinfo {author} {\bibfnamefont {S.~S.}\ \bibnamefont {Ray}},\ }\href
  {https://doi.org/10.1103/PhysRevLett.132.184002} {\bibfield  {journal}
  {\bibinfo  {journal} {Phys. Rev. Lett.}\ }\textbf {\bibinfo {volume} {132}},\
  \bibinfo {pages} {184002} (\bibinfo {year} {2024})}\BibitemShut {NoStop}%
\bibitem [{\citenamefont {Ray}(2018)}]{Ray-Irreversible}%
  \BibitemOpen
  \bibfield  {author} {\bibinfo {author} {\bibfnamefont {S.~S.}\ \bibnamefont
  {Ray}},\ }\href {https://doi.org/10.1103/PhysRevFluids.3.072601} {\bibfield
  {journal} {\bibinfo  {journal} {Phys. Rev. Fluids}\ }\textbf {\bibinfo
  {volume} {3}},\ \bibinfo {pages} {072601} (\bibinfo {year}
  {2018})}\BibitemShut {NoStop}%
\end{thebibliography}%
\appendix
\onecolumngrid 

\section{Appendix A: Scaling form for the second-order structure functions} 
\renewcommand{\theequation}{A-\arabic{equation}}
\setcounter{equation}{0} 

\begin{figure}
	\includegraphics[width = 1.0\linewidth]{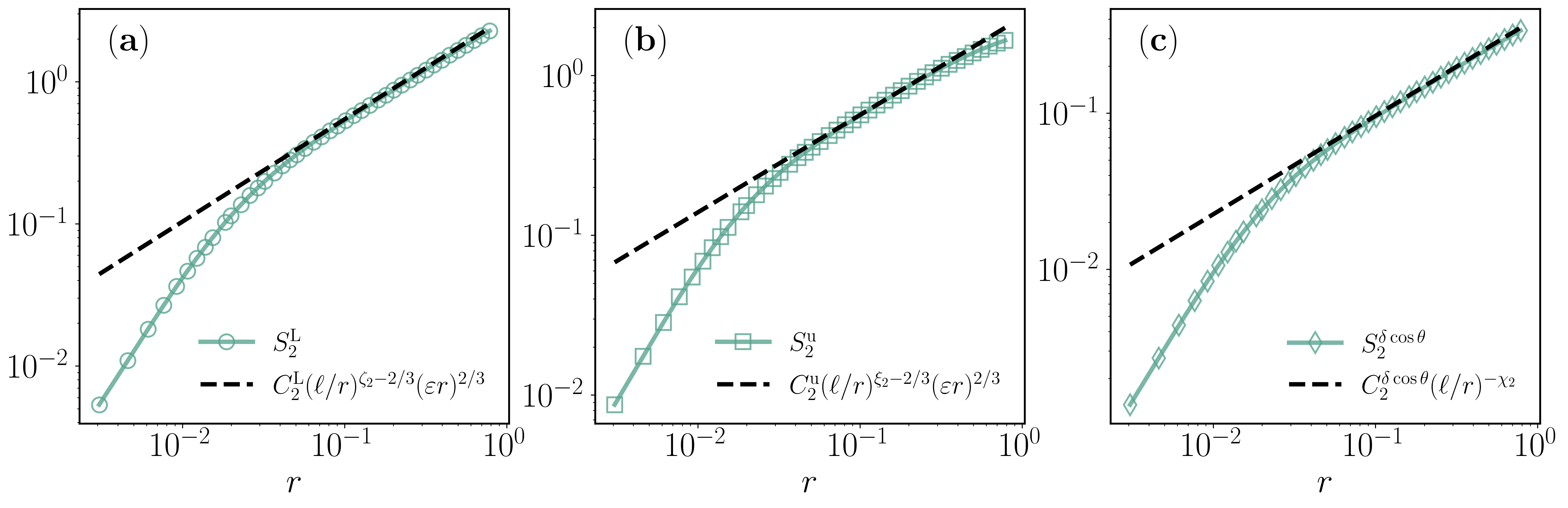}
	\caption{Loglog plots of the different second-order structure functions for increments of (a) the longitudinal velocity (b) 
	velocity amplitudes and (c) the cosine-angles. Each panel is accompanied by the scaling form (valid in the inertial range) 
	from which factors in the anomalous part of the scaling; for (c) we do not have a dimensional form for the 
	scaling since the cosine-angle structure functions are dimensionaless.}
	\label{fig:ScalingS2}
\end{figure}

We show the accuracy of the scaling form (see main text)
\begin{eqnarray}
	S^{\rm L}_2(r) &=& C_2^{\rm L} \left (\frac{\ell}{r}\right )^{p/3 - \zeta^{\rm L}_p}\left (\varepsilon r \right) ^{p/3}\\
	S^{\rm u}_2(r) &=& C_2^{\rm u} \left (\frac{\ell}{r}\right )^{p/3 - \xi_p}\left (\varepsilon r \right) ^{p/3}\\
	S^{\delta \cos \theta}_2(r) &=&  C_2^{\delta \cos \theta}  \left (\frac{\ell}{r}\right )^{- \chi_p}
\label{eq:ScalingS2}
\end{eqnarray}
of the second-order structure functions. Of course, such a form for longitudinal increments is well known~\cite{Frisch-Book}, 
and in Fig.~\ref{fig:ScalingS2}(a) we fit this form to extract $C_2^{\rm L} = 2.3$~\cite{Ni2013} by using the integral length scale $\ell = 1.39$ and mean dissipation rate $\varepsilon = 1.41$. 
We then use the same $L$ to fit, in panels (b) and (c), respectively, the second-order structure function $S_2^{\rm u}$ for the 
amplitude and $S_2^{\delta \cos \theta}$ for the cosine-angle increments. In particular, our fits yield $C^{\rm u} \approx 1.8$ and 
$C^{\delta \cos \theta} \approx 0.5$.

\section{Appendix B: Multiscaling Exponents and the Probability Distribution of Increments} 
\renewcommand{\theequation}{B-\arabic{equation}}
\setcounter{equation}{0} 

In this section we analyse the multiscaling exponents for the $p$-order, equal-time structure function exponents, associated with the 
increments $\langle |\delta \sin \theta_r|^p \rangle \equiv \langle |\sin \theta_{\rm A} - \sin \theta_{\rm B}|^p \rangle 
\sim r^{\beta_p}$, $\langle |\sin \delta \theta_r |^p\rangle \equiv \langle |\sin [\theta_{\rm A} - \theta_{\rm B}]|^p \rangle \sim r^{\gamma_p}$, and 
$\langle |\delta \theta_r |^p \rangle \equiv \langle |[\theta_{\rm A} - \theta_{\rm B}]|^p\rangle \sim r^{\eta_p}$. 

\begin{figure}
	\includegraphics[width = 0.32\linewidth]{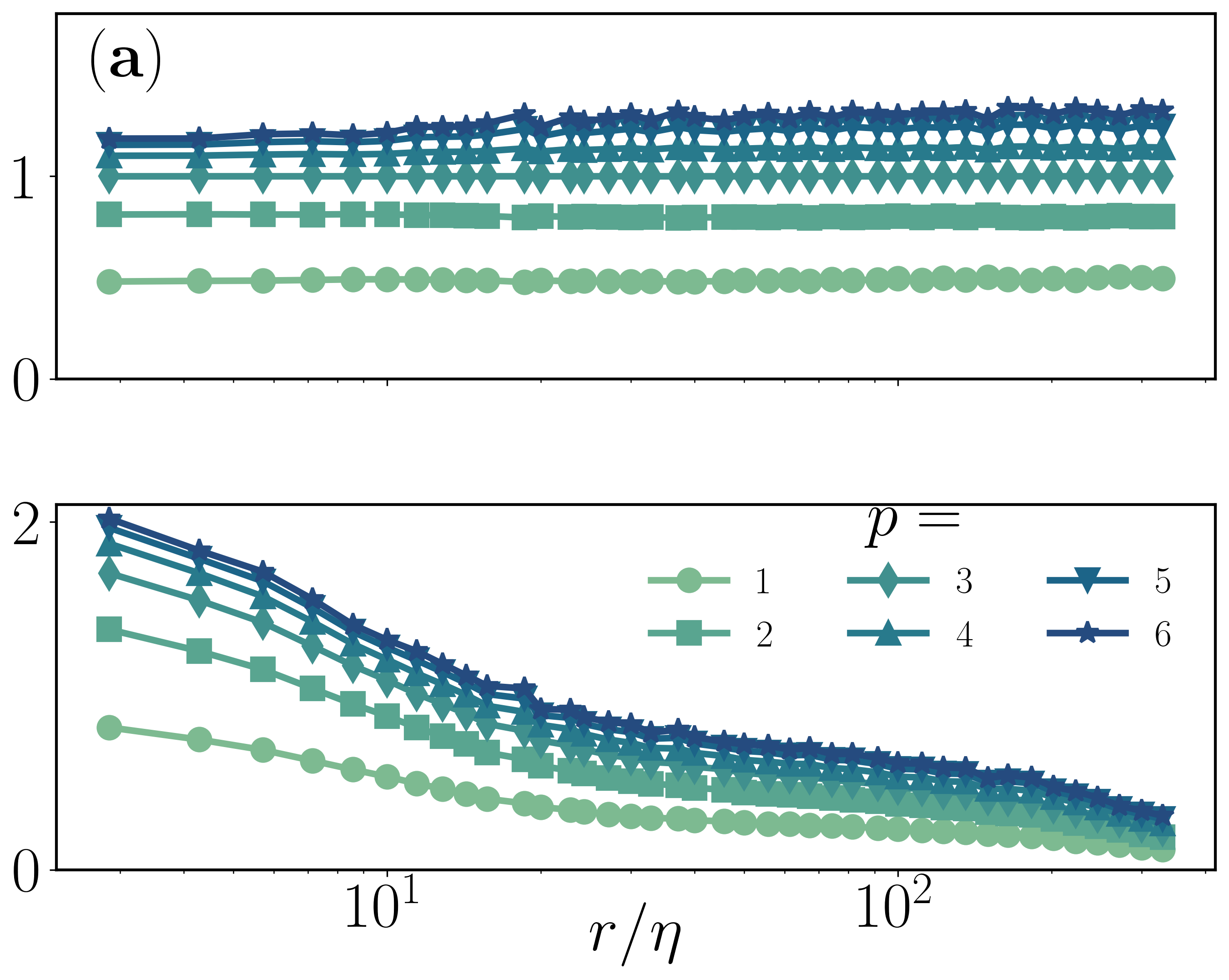}
	\includegraphics[width = 0.32\linewidth]{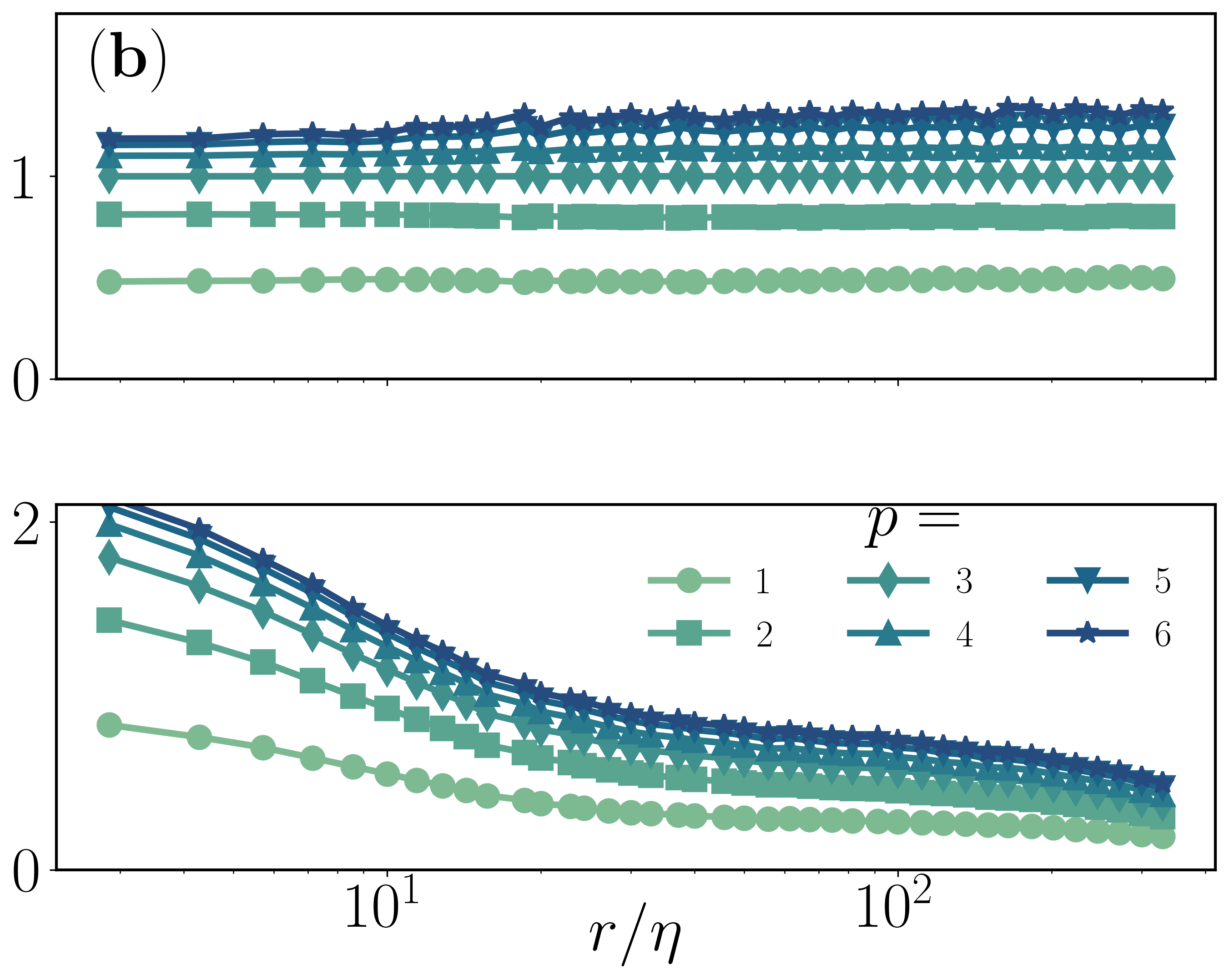}
	\includegraphics[width = 0.32\linewidth]{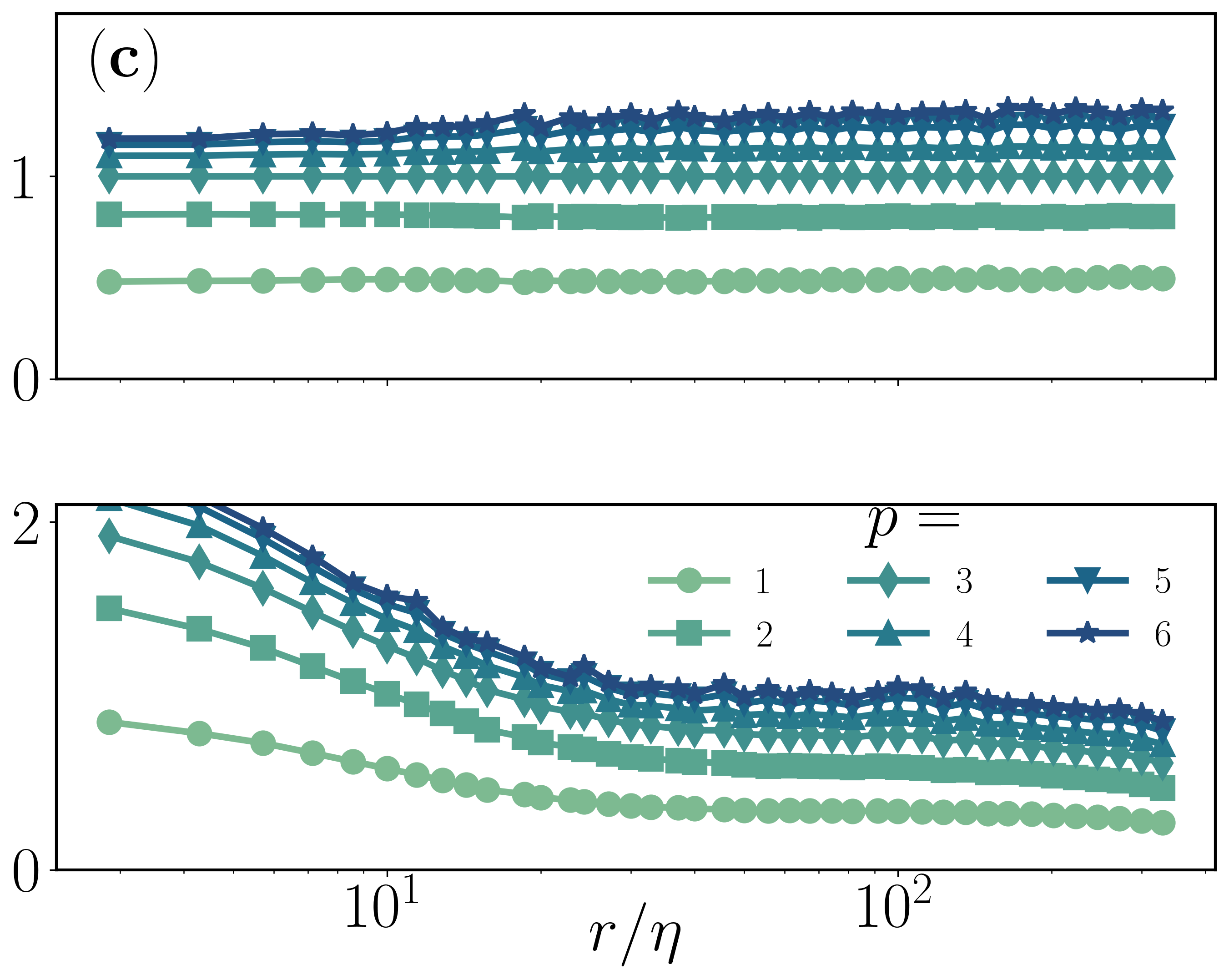}
	\caption{Plots of the local slopes for structure functions increments   
	$\delta\sin\theta_r$ (b) $\delta\sin\theta_r$ and (c) $\delta\theta_r$ with (upper panel) and without (lower panel) 
	the use of extended self-similarity in three-dimensional turbulence. A similar quality of scaling 
	is also obtained for two-dimensional turbulence.}
	\label{fig:AppA-zetaP}
\end{figure}

These exponents are obtained through a local slope analysis (as discussed in
the main text; see Fig.~\ref{fig:AppA-zetaP}). The bare exponents, as well as
those obtained via ESS, are listed in Table~\ref{tab:add_exponents_3D}, for
three-dimensional and in Table~\ref{tab:add_exponents_2D}, for two-dimensional
turbulence.

\begin{figure}
	\includegraphics[width = 1.0\linewidth]{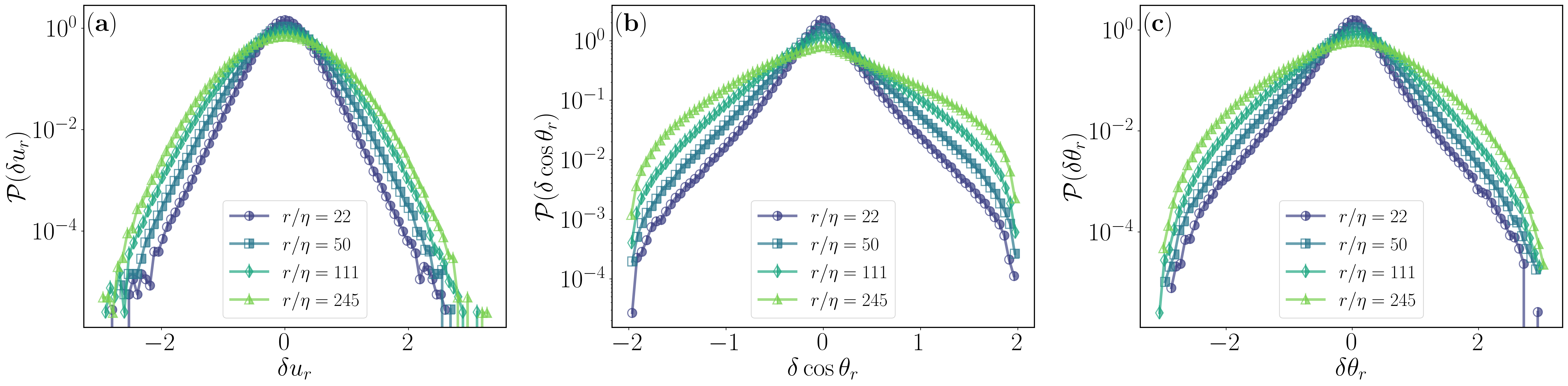}
	\caption{Representative plots of the probability distribution functions of (a) $\delta u_r$, (b) $\delta \cos \theta_r$, and (c) $\delta \theta_r$ 
for different values of $r/\eta$.}
	\label{fig:AppA-PDF}
\end{figure}

Multiscaling of inertial range exponents are associated with non-Gaussian tails of the probability distribution functions of the increments themselves. In 
Fig.~\ref{fig:AppA-PDF} we show representative plots of the probability distribution functions of (a) $\delta u_r$, (b) $\delta \cos \theta_r$, and (c) $\delta \theta_r$ 
for different values of $r/\eta$, which show clear non-Gaussian tails. In particular, for the velocity amplitude increments (Fig.~\ref{fig:AppA-PDF}(a)), large 
values of $r/\eta$ lead to a Gaussian distribution with ever widening tails as $r/\eta$ fall in the inertial range. 

\begin{table}
	\centering
	\begin{tabular}{|c|c|c|c|c|c|c|}
		\hline
		$p$ & $\beta_p$ & $\tilde{\beta}_p$ & $\gamma_p$ & $\tilde{\gamma}_p$ & $\eta_p$ & $\tilde{\eta}_p$ \\
		\hline
		1 & $0.30 \pm 0.03$ & $0.462 \pm 0.005$ & $0.32 \pm 0.02$ & $0.457 \pm 0.005$ & $0.36 \pm 0.01$ & $0.424 \pm 0.004$ \\
		2 & $0.47 \pm 0.04$ & $0.790 \pm 0.003$ & $0.51 \pm 0.03$ & $0.786 \pm 0.003$ & $0.61 \pm 0.02$ & $0.769 \pm 0.005$ \\
		3 & $0.56 \pm 0.06$ & $1.00$ & $0.63 \pm 0.04$ & $1.00$ & $0.79 \pm 0.02$ & $1.00$ \\
		4 & $0.63 \pm 0.06$ & $1.128 \pm 0.006$ & $0.72 \pm 0.05$ & $1.131 \pm 0.005$ & $0.92 \pm 0.02$ & $1.15 \pm 0.01$ \\
		5 & $0.68 \pm 0.07$ & $1.22 \pm 0.01$ & $0.77 \pm 0.05$ & $1.22 \pm 0.01$ & $0.98 \pm 0.02$ & $1.24 \pm 0.03$ \\
		6 & $0.72 \pm 0.07$ & $1.28 \pm 0.02$ & $0.81 \pm 0.05$ & $1.28 \pm 0.02$ & $1.03 \pm 0.02$ & $1.28 \pm 0.04$ \\
		\hline
\end{tabular}
\caption{A summary of the equal-time scaling exponents $\beta_p$, $\gamma_p$, and $\eta_p$, with the corresponding exponent ratios obtained via ESS 
(denoted by a \textit{tilde}) for three-dimensional turbulence.} 
\label{tab:add_exponents_3D}
\end{table}

\begin{table*}
	\centering
\begin{tabular}{|c|c|c|c|c|c|c|}
	\hline
	$p$ & $\beta_p$ & $\tilde{\beta}_p$ & $\gamma_p$ & $\tilde{\gamma}_p$ & $\eta_p$ & $\tilde{\eta}_p$ \\
	\hline
	1 & $0.23 \pm 0.04$ & $0.55 \pm 0.03$ & $0.21 \pm 0.04$ & $0.54 \pm 0.02$ & $0.30 \pm 0.03$ & $0.49 \pm 0.03$ \\
	2 & $0.34 \pm 0.06$ & $0.83 \pm 0.01$ & $0.32 \pm 0.06$ & $0.83 \pm 0.01$ & $0.49 \pm 0.04$ & $0.80 \pm 0.03$ \\
	3 & $0.41 \pm 0.07$ & $1.00 $ & $0.39 \pm 0.07$ & $1.00 $ & $0.59 \pm 0.03$ & $1.00 $ \\
	4 & $0.45 \pm 0.07$ & $1.11 \pm 0.01$ & $0.43 \pm 0.08$ & $1.11 \pm 0.01$ & $0.65 \pm 0.01$ & $1.10 \pm 0.03$ \\
	5 & $0.47 \pm 0.08$ & $1.18 \pm 0.03$ & $0.46 \pm 0.08$ & $1.19 \pm 0.02$ & $0.69 \pm 0.01$ & $1.16 \pm 0.05$ \\
	6 & $0.49 \pm 0.08$ & $1.24 \pm 0.04$ & $0.48 \pm 0.08$ & $1.24 \pm 0.02$ & $0.71 \pm 0.02$ & $1.21 \pm 0.07$ \\
	\hline
\end{tabular}
\caption{A summary of the equal-time scaling exponents $\beta_p$, $\gamma_p$, and $\eta_p$, with the corresponding exponent ratios obtained via ESS 
(denoted by a \textit{tilde}) for two-dimensional turbulence.} 
\label{tab:add_exponents_2D}
\end{table*}

\end{document}